\documentclass[pre,twocolumn,showpacs]{revtex4}
\usepackage{amsmath}

\begin{document}

\title{Nonlinear interfacial waves in a constant-vorticity planar flow over variable depth
}
\author{V. P. Ruban}
\email{ruban@itp.ac.ru}
\affiliation{Landau Institute for Theoretical Physics,
2 Kosygin Street, 119334 Moscow, Russia} 

\date{\today}

\begin{abstract}
Exact Lagrangian in compact form is derived for planar internal waves in a two-fluid system 
with a relatively small density jump (the Boussinesq limit taking place in real oceanic
conditions), in the presence of a background  shear current of constant vorticity, 
and over arbitrary bottom profile. Long-wave asymptotic approximations of higher orders 
are derived from the exact Hamiltonian functional in a remarkably simple way, for two
different parametrizations of the interface shape.
\end{abstract}

%47.35.-i       Hydrodynamic waves
%47.10.Df 	Hamiltonian formulations 
%47.55.-t 	Multiphase and stratified flows
%47.15.K- 	Inviscid laminar flows 
%%%%%%%%%%%%%%%%%%%%%%%%%%%%%%%%%%%%%%%%%%%%%%%%%%%%%%%%%%%%%%%%%%%%%%%%%%%

\maketitle

Large-amplitude internal waves are important phenomena in near-coastal ocean dynamics (see, e.g.,
Refs.\cite{AHLT_1985,LHA_1985,BRAB_1997,DR_1978,VBR_2000,VH_2002,GPTK_2004,VS_2006},
and references therein).
A number of theoretical models is used to study internal waves analytically, including the most 
popular two-fluid model \cite{B_1966,B_1967,O_1975,J_1977}, where the system is placed between
fixed upper and bottom boundaries and consists of two fluid layers having different constant
densities ($\rho_1=1$ in the upper layer, and $\rho_2=1+\epsilon$ in the lower layer).
In the simplest variant, the rigid walls are both horizontal, and the flow is potential
within each layer, but additional ingredients can be incorporated into the model, 
such as a background shear current and large bathymetric variations
(see Refs.\cite{BBVG_1993,SL_2002,C_2006,deZN_2008,deZVNC_2009}, and references therein).
In the general case, an analytical study of this system requires quite lengthy calculations, 
especially when fully nonlinear dispersive approximations of higher orders are considered
\cite{deZN_2008,deZVNC_2009,CC_1999,OG_2003,CGK_2005,CCMRS_2006,BLS_2008,CBJ_2009,DDK_2010}. 
In the present work, it will be shown that in the case $\epsilon\ll 1$
(the so called Boussinesq limit), there exists an elegant and remarkably 
short way how to derive fully nonlinear dispersive models of high orders. 
It should be noted that in real oceanic circumstances $\epsilon$ is indeed small, 
typically $\epsilon\sim 0.01$, and therefore the theory developed below is of practical
importance. Moreover, our approach easily takes into account the presence of a shear 
background current of constant vorticity, as well as arbitrary variations of the bed profile, 
and that makes the model more adequate for studying the real-world  processes of interaction 
of internal waves with tidal currents over non-uniform sea bed. We also concern in this work
the problem of proper parametrization of steep wave profiles and, besides the usual 
single-valued $y=\eta(x,t)$ representation, we consider an alternative parametrization 
which is more suitable to describe arbitrary steep and even multi-valued wave profiles.

\emph{Variational formalism. ---}
The method employed here is based on a variational formulation for the interface dynamics
in a two-fluid planar system having constant vorticities $\Omega_{1,2}$ within each layer.
Let the (unknown) interface profile be $y=\eta(x,t)$, where $y$ is the vertical coordinate,
$x$ is the horizontal coordinate, and $t$ is the time variable. 
A two-dimensional (2D) velocity field can be represented as follows 
(subscripts $1,2$ numbering the layers are omitted),
\begin{equation}
{\bf v}(x,y,t)=\Omega {\bf S}(x,y) +\nabla\varphi(x,y,t),
\end{equation}
where the stationary divergence-free field ${\bf S}(x,y)$ satisfies equation 
$\mbox{curl }{\bf S}=1$ and has zero normal component at the rigid boundaries.
The unknown potentials $\varphi_{1,2}(x,y,t)$ satisfy the Laplace equation
$(\partial_x^2+\partial_y^2)\varphi_{1,2}=0$, with the boundary conditions 
$(\nabla\varphi_{1,2}\cdot{\bf n})=0$ at a rigid wall (where ${\bf n}$ the normal 
unit vector), and $\varphi_{1,2}[x,\eta(x,t),t]=\psi_{1,2}(x,t)$ at the interface.
Since there is a kinematic boundary condition at the interface,
\begin{equation}
({\bf v}_1\cdot {\bf n})=({\bf v}_2\cdot {\bf n})=V_n=\eta_t/\sqrt{1+\eta_x^2},
\end{equation}
functions $\psi_{1}(x,t)$ and $\psi_{2}(x,t)$ are related to each other by 
a linear integral operator depending on $\eta(x,t)$. 
Therefore there are only two basic unknown functions, $\eta(x,t)$ and, for instance, 
\begin{equation}
\psi(x,t)=\psi_1(x,t)-(1+\epsilon)\psi_2(x,t).
\end{equation}
It can be proved that the Lagrangian functional for the system under consideration
has the following structure,
\begin{equation}\label{L_gamma}
{\cal L}_{general}=-\int\psi\eta_t dx +\frac{\gamma}{2}\int\eta\partial_x^{-1}\eta_t dx 
-{\cal H}\{\eta,\psi\},
\end{equation}
where $\gamma=(1+\epsilon)\Omega_2-\Omega_1$, and the Hamiltonian functional is
the total energy of the system --- kinetic plus potential. 
The general idea of the proof is similar to Refs.\cite{W_2007,W_2008,R2010JETP}, 
and it is based on a generalization of the Bernoulli equation 
for constant-vorticity 2D incompressible flows (subscripts $1,2$ are omitted),
\begin{equation}
\partial_t\varphi+\Omega\Theta+{\bf v}^2/2+g y +P/\rho=0,
\end{equation}
where $\Theta(x,y,t)$ is the total stream function in the layer, 
$g$ is the gravity acceleration, and $P(x,y,t)$ is the pressure.
At the interface, $\Theta$ and $P$ should be continuous, and this requirement gives us
the dynamic boundary condition 
\begin{equation}\label{dyn_boundary_cond}
\Big[(1+\epsilon)\partial_t\varphi_2-\partial_t\varphi_1 +\gamma\Theta
+(1+\epsilon)\frac{{\bf v}_2^2}{2}-\frac{{\bf v}_1^2}{2}\Big]\Big|_{y=\eta}+\epsilon g \eta=0.
\end{equation}
Since, by definition, the relation
\begin{equation}
\partial_t\psi_{1,2}=[\partial_t\varphi_{1,2}+\eta_t \partial_y\varphi_{1,2}]|_{y=\eta}
\end{equation}
takes place, and $\eta_t=\partial_x(\Theta|_{y=\eta})$, Eq.(\ref{dyn_boundary_cond}) 
is equivalent to 
\begin{equation}\label{Ham_dynamic}
\psi_t+\gamma\partial_x^{-1}\eta_t=\delta{\cal H}/\delta\eta,
\end{equation}
while the kinematic boundary condition can be represented as
\begin{equation}\label{Ham_kinematic}
-\eta_t=\delta{\cal H}/\delta\psi. 
\end{equation}
Equations (\ref{Ham_dynamic}) and (\ref{Ham_kinematic}) are equivalent to Eq.(\ref{L_gamma}).
Let us for example prove that $V_n\sqrt{1+\eta_x^2}=-\delta{\cal H}/\delta\psi$ 
(consideration of the dynamic boundary condition is slightly more involved but analogous). 
Since $\psi$ is present in the kinetic energy only, the corresponding variation of the 
Hamiltonian is
\begin{eqnarray}
\delta{\cal H}&=&\!\int_{D1} \!({\bf v}_1\cdot\nabla\delta\varphi_1) dx dy 
+(1+\epsilon)\!\int_{D2}\! ({\bf v}_2\cdot\nabla \delta\varphi_2) dx dy\nonumber\\
&=&\int V_n\sqrt{1+\eta_x^2}[(1+\epsilon)\delta\psi_2-\delta\psi_1]dx,
\end{eqnarray}
where we have integrated by parts and used the boundary conditions
and incompressibility. Taking into account the definition of $\psi$, 
we arrive at the required result.

\emph{Exact Hamiltonian theory in Boussinesq limit. ---}
The main technical difficulty for application of the Hamiltonian formalism at finite $\epsilon$ 
is the absence of compact expression for the kinetic energy of the flow 
in terms of functions $\eta$ and $\psi$. Therefore,
below we consider the case $\epsilon\ll 1$ and, for simplicity, $\Omega_1=\Omega_2=\Omega$, 
resulting in $\gamma\approx 0$. 
What is very essential, in this limit $\psi\approx\psi_1-\psi_2$, which means
that $-\psi_x(x,t)dx$ is the strength of singular vorticity concentrated at the interface. 
Since the kinetic energy of the system in the Boussinesq limit is given by integral
$
{\cal K}_{total}\approx(1/2)\int\Theta\Omega_{total}dxdy, 
$
and 
\begin{equation}
(\partial_x^2+\partial_y^2)\Theta=-\Omega_{total}=-\Omega+\psi_x(x,t)\delta[y-\eta(x,t)],
\end{equation}
the above observation allows us to express the kinetic energy in a closed 
form through the Green's function of the 2D Laplace operator in the domain 
between the fixed boundaries, with zero boundary conditions.
An explicit expression for the Green's function is known in terms of
curvilinear conformal coordinates $u$ and $v$, with $v=0$ at the upper boundary, $v=-1$
at the bottom, $(\partial_x^2+\partial_y^2)v=0$ inside the domain,
and $u(x,y)$ being a harmonically conjugate for $v(x,y)$ \cite{R2004PRE,R2010PRE}.
In other words, $x+iy=z(u+iv)$, where $z(w)$ is an analytic function of complex variable
$w=u+iv$. We can write the Lagrangian in terms of the interface profile $v=-q(u,t)$ 
and $\psi(u,t)$,
\begin{equation}
{\cal L}=\int\psi J(u,q)q_t dx -{\cal H}\{\psi,q\} ,
\end{equation}
where $J(u,q)=|z'(u-iq)|^2$ is the Jacobian. The corresponding equations of motion are
\begin{equation}\label{Ham_eqs}
J(u,q)q_t=\delta{\cal H}/\delta\psi,\qquad -J(u,q)\psi_t=\delta{\cal H}/\delta q.
\end{equation}
The Hamiltonian consists of three parts, ${\cal H}={\cal P}+{\cal K}+{\cal S}$.
The potential energy is
\begin{equation}\label{P}
{\cal P}=\frac{\epsilon g}{2}\int[\mbox{Im }z(u-iq)]^2\mbox{Re}[\partial_uz(u-iq)]du.
\end{equation}
${\cal K}$ is the kinetic energy in the absence of external current,
\begin{eqnarray}
{\cal K}&=&\frac{1}{4\pi}\iint\ln\Big|\frac{\sinh[(\pi/2)(u_2-u_1-i(q_2+q_1))]}
{\sinh[(\pi/2)(u_2-u_1-i(q_2-q_1))]}\Big|\nonumber\\
&&\qquad\qquad\times\psi'_1\psi'_2d u_1 du_2,
\end{eqnarray}
where $q_1=q(u_1,t)$, $\psi'_1=[\partial_u\psi(u,t)]|_{u=u_1}$, and so on.
The functional ${\cal S}$ takes into account the presence of a background shear current,
\begin{equation}\label{S}
{\cal S}=\int[Cq-\Omega\Theta_s(u,-q)]\psi' du +\mbox{const}.
\end{equation}
Here $C$ is a total flux of the current, and $\Theta_s(u,v)$ is the function 
which satisfies zero boundary conditions and equation $(\partial_x^2+\partial_y^2)\Theta_s=-1$. 
Usually it is assumed that the upper rigid boundary is 
horizontal at $y=0$, and that results in $y(u,-v)=-y(u,v)$. Then the stream function 
$\Theta_s(u,v)$ can be represented as follows (the idea how to calculate $\Theta_s$ 
is similar to Ref.\cite{R2008PRE}),
\begin{equation}\label{Theta_s}
\Theta_s(u,v)=-\frac{1}{2}\Big\{y^2(u,v)+
\Big[\frac{\sinh\hat k v}{\sinh\hat k}\Big]y^2(u,-1)\Big\},
\end{equation}
where $\hat k=-i\partial_u$ is the differential operator, which is diagonal in Fourier
representation.

Let us say here that the case $\gamma\not= 0$, though more cumbersome,  
can be considered in a similar manner.

\emph{Long-wave approximations. ---}
Thus, all the terms in exact Hamiltonian functional have been explicitly specified.
However, this exact description cannot be applied easily for analytical and numerical studies 
in view of the strongly non-local character of the Hamiltonian. This situation with internal
waves is in contrast with the exact description of 2D surface waves of constant vorticity
\cite{R2008PRE}, where at least numerical implementation is simple and efficient with 
a fast Fourier transform. That is why the problem of simplified description for 
moderately steep interfacial waves has attracted much attention in last years
\cite{deZN_2008,deZVNC_2009,CC_1999,OG_2003,CGK_2005,CCMRS_2006,BLS_2008,CBJ_2009,DDK_2010}.
Here we suggest a simple procedure for the Boussinesq case.
We see that the only non-local part in the total Hamiltonian is ${\cal K}$. 
To derive simplified quasi-local approximations, we will essentially use a Fourier transform
of the Green's function,
\begin{equation}
\frac{1}{2\pi}\ln\Big|\frac{\sinh[(\pi/2)(\tilde u-is)]}
{\sinh[(\pi/2)(\tilde u-ia)]}\Big|=\int F(k,a,s)e^{ik\tilde u}\frac{dk}{2\pi},
\end{equation}
where $\tilde u=u_2-u_1$, $a=|q_2-q_1|$, $s=q_2+q_1$, and
\begin{equation}\label{F_kas}
F(k,a,s)=\frac{\cosh k(1-a)-\cosh k(1-s)}{2k\sinh k}.
\end{equation}
This expression can be easily obtained by the known methods of contour integration 
of analytic functions.
It is the simplicity of Eq.(\ref{F_kas}) that allows us to derive the mentioned
dispersive models of high orders for nonlinear internal waves. The approximations correspond 
to expansion  of $F$ in powers of small quantities $ka$ and $ks$. Therefore, the long-wave
assumption actually includes three different asymptotic regimes: 
(i) deep-water theory, when $k\gg 1$ but $kq\ll 1$,
(ii) finite-depth theory, when $q\ll 1$ and $k\sim 1$, 
(iii) shallow-water theory, when $q\sim 1$ and $k\ll 1$ [for self-consistency, in this regime
the bottom variations should be long-scaled as well, $|x''(u)|/x'(u)\ll 1$,
where $x(u)=z(u+0i)$ is a purely real function]. Thus, we write
\begin{eqnarray}
F&=&\frac{s-a}{2}+\frac{a^2-s^2}{4}k\coth k+\cdots\nonumber\\
&+&\frac{(s^{2n-1}-a^{2n-1})}{2(2n-1)!}k^{2n-2}\nonumber\\
&+&\frac{(a^{2n}-s^{2n})}{2(2n)!}k^{2n-1}\coth k +\cdots.
\end{eqnarray}
It is not difficult to understand that each term proportional to $a^{2m+1}k^{2m}$
gives zero contribution to the Hamiltonian. The other terms are easily transformed to
finite sums of products as $\psi'_1q_1^{m_1}\psi'_2q_2^{m_2}$, thus resulting in the 
quasi-local expansion ${\cal K}={\cal K}_1+{\cal K}_2+{\cal K}_3+\cdots$,  where
\begin{eqnarray}
{\cal K}_1&=&\frac{1}{2}\int q\psi'^2 du,\\
{\cal K}_2&=&-\frac{1}{2}\int\psi'_1 q_1\psi'_2 q_2 [k\coth k]e^{ik(u_2-u_1)}
\frac{dk}{2\pi}du_1du_2
\nonumber\\
&=&-\frac{1}{2}\int q\psi'[\hat k\coth\hat k](q\psi')du,
\end{eqnarray}
\begin{eqnarray}
{\cal K}_3&=&\frac{1}{4\cdot3!}\int(q_1+q_2)^3\psi'_1\psi'_2 e^{ik(u_2-u_1)}
\frac{dk}{2\pi}du_1du_2
\nonumber\\
&=&\frac{1}{12}\int [\psi''(q^3\psi')' +3(q^2\psi')'(q\psi')'] du,
\end{eqnarray}
\begin{eqnarray}
{\cal K}_4&=&\frac{1}{4\cdot4!}\int [(q_1-q_2)^4-(q_1+q_2)^4]\psi'_1\psi'_2\nonumber\\
&&\qquad\qquad\times [k^3\coth k]e^{ik(u_2-u_1)}\frac{dk}{2\pi}du_1du_2
\nonumber\\
&=&-\frac{1}{6}\int (q\psi')'[\hat k\coth\hat k](q^3\psi')' du.
\end{eqnarray}
Analogously, we derive
\begin{eqnarray}
{\cal K}_5&=&\frac{1}{240}\int\Big[(q^5\psi')''\psi'''+5(q^4\psi')''(q\psi')''\nonumber\\
&&\qquad\qquad+10(q^3\psi')''(q^2\psi')''\Big]du,\\
{\cal K}_6&=&-\frac{1}{360}\int\Big[3(q^5\psi')'' [\hat k\coth\hat k](q\psi')''\nonumber\\
&&\qquad\qquad+10(q^3\psi')'' [\hat k\coth\hat k](q^3\psi')''\Big]du,\\
{\cal K}_7&=&\frac{1}{2\cdot 7!}\int\Big[(q^7\psi')'''\psi^{(4)}
+7(q^6\psi')'''(q\psi')'''\nonumber\\
&+&21(q^5\psi')'''(q^2\psi')'''\!+\!35(q^4\psi')'''(q^3\psi')'''\Big]du.
\end{eqnarray}
It is a simple exercise to calculate the variational derivatives of 
the obtained approximate Hamiltonian and substitute them into Eqs.(\ref{Ham_eqs}).
Hopefully, with appropriate regularization at short scales, 
the equations will appear to be convenient for future numerical implementation, 
since all the linear operators 
(the differentiations and $[\hat k\coth\hat k]$) are diagonal in Fourier representation.

The functionals ${\cal P}$ and ${\cal S}$ take a simplified form in the case of 
a slowly varying depth, when the series
\begin{equation}\label{z_expansion}
z(u-iq)=x(u)-i x'(u)q-x''(u)\frac{q^2}{2}+ix'''(u)\frac{q^3}{6} +\cdots
\end{equation}
rapidly converges. Then substitution of Eq.(\ref{z_expansion}) into Eqs.(\ref{P}) and
(\ref{Theta_s}) gives us
\begin{equation}
{\cal P}\approx \frac{\epsilon g}{2}\int 
\Big\{q^2x'^3+q^4\Big[\frac{(x'^2x'')'}{4}-\frac{5}{6}x'^2x'''\Big]\Big\}du,
\end{equation}
\begin{eqnarray}
{\cal S}&\approx& \int \psi'
\Big\{Cq+\frac{\Omega}{2}\Big[q^2\Big(x'^2-\frac{x'x'''}{3}q^2\Big)\nonumber\\
&&-q\Big(x'^2-\frac{x'x'''}{3}\Big)+
\frac{(x'^2)''}{6}\Big(q^3-q\Big)\Big]\Big\}du.
\end{eqnarray}
Note that the Jacobian is this case is
\begin{equation}
J(u,q)\approx x'^2+[(x'')^2-x'x''']q^2. 
\end{equation}

In the shallow-water regime $k\coth k\approx1+k^2/3$, and ${\cal K}$ takes a purely local form,
\begin{eqnarray}
{\cal K}_{local}&=&\frac{1}{2}\int q(1-q)\psi'^2 du\nonumber\\
&+&\frac{1}{12}\int\left\{(q\psi')'[(1-q)\psi']'-[q(1-q)\psi']\psi''\right\}du\nonumber\\
&-&\frac{1}{12}\int[q^2(1-q)\psi']'[(1-q)\psi']'du\nonumber\\
&-&\frac{1}{12}\int[(1-q)^2q\psi']'(q\psi')'du +{\cal O}\{\partial_u^6\},
\end{eqnarray}
which is apparently symmetric with respect to change $q\to (1-q)$. This symmetry is present also
in the exact expression (\ref{F_kas}), since if $a\to a$ and $s\to (2-s)$, then $F\to F$.

\emph{Alternative parametrization. ---}
Strongly nonlinear internal waves are known to have the tendency toward overturning their
profiles. The above theory can be generalized to admit more steep, and even multi-valued
dependences $y=\eta(x,t)$. To explain the basic idea how to manage in that case, 
we consider here the deep-water regime in the absence of current. 
An arbitrary interface shape can be represented in a parametric form,
\begin{equation}
x=X(\sigma,t)\equiv\sigma +\tilde X(\sigma,t),\qquad y=Y(\sigma,t)<0, 
\end{equation}
with a parameter $-\infty<\sigma<+\infty$ along the curve. What is important,
smooth functions $\tilde X(\sigma,t)$ and $Y(\sigma,t)$ are able to represent quite steep
wave profiles. The corresponding  Lagrangian is
\begin{eqnarray}
&&{\cal L}_{param}=\int[X' Y_t-X_t Y']\psi d\sigma-
\frac{\epsilon g}{2}\int Y^2X' d\sigma\nonumber\\
&&-\frac{1}{8\pi}\!\int\!\ln\Bigg[\frac{(X_2\!-\!X_1)^2\!+\!(Y_2\!+\!Y_1)^2}
{(X_2\!\!-X_1)^2\!+\!(Y_2\!-\!Y_1)^2}\Bigg]\psi'_1\psi'_2 d\sigma_1 d\sigma_2.
\end{eqnarray}
To simplify the non-local term $\tilde{\cal K}$, 
we again use a Fourier transform of the Green's function, 
but in a slightly different manner,
\begin{eqnarray}
&&\frac{1}{4\pi}\ln\Bigg[\frac{(\sigma_2-\sigma_1+\tilde X_2-\tilde X_1)^2+(Y_2+Y_1)^2}
{(\sigma_2-\sigma_1+\tilde X_2-\tilde X_1)^2+(Y_2-Y_1)^2}\Bigg]\nonumber\\
&&=\int\tilde F(\kappa,\tilde X_2,Y_2,\tilde X_1,Y_1)
 e^{i\kappa(\sigma_2-\sigma_1)}\frac{d\kappa}{2\pi},
\end{eqnarray}
with [note the appearance of factor $\exp\{i\kappa(\tilde X_2-\tilde X_1)\}$]
\begin{equation}\label{tilde_F}
\tilde F=e^{i\kappa(\tilde X_2-\tilde X_1)}
\Big[\frac{e^{-|\kappa||Y_2-Y_1|}-e^{-|\kappa||Y_2+Y_1|}}{2|\kappa|}\Big].
\end{equation}
Now we expand the exponents in Eq.(\ref{tilde_F}) in powers of the arguments, 
and take into account that terms proportional to 
$(\tilde X_2-\tilde X_1)^m|Y_2-Y_1|^{2n+1}k^{m+2n}$ 
give zero contribution to the Hamiltonian. As the result, we obtain
\begin{eqnarray}
\tilde {\cal K}&=&-\frac{1}{2}\int Y\psi'^2d\sigma 
-\frac{1}{2}\int(Y\psi')|\hat\kappa|( Y\psi')d\sigma\nonumber\\
&&-\frac{1}{2}\int[\tilde X Y\psi'\psi''+\tilde X \psi'(Y\psi')']d\sigma\nonumber\\
&&-\frac{1}{12}\int[\psi''(Y^3\psi')' +3(Y^2\psi')'(Y\psi')']d\sigma\nonumber\\
&&+\frac{1}{4}\int\Big[(Y\tilde X^2\psi')'\psi''
-2(Y\tilde X\psi')'(\tilde X\psi')'\nonumber\\
&&\qquad\qquad\qquad +(Y\psi')'(\tilde X^2\psi')'\Big]d\sigma\nonumber\\
&&+\int Y\psi'|\hat\kappa|(\tilde X Y\psi')'d\sigma+{\cal O}\{\partial_\sigma^5\},
\label{tilde_K}
\end{eqnarray}
where $|\hat\kappa|=|\partial_\sigma|$ is a pseudo-differential operator.

It should be noted that if the parametrization of interface is arbitrary, 
without any relation between $\tilde X(\sigma,t)$ and $Y(\sigma,t)$, 
then the tangential component of the interface motion (the combination $X'X_t+Y'Y_t$) 
is not determined by variational equations of motion, and therefore it remains arbitrary.
One can somehow fix the parametrization, for example by relation 
$\tilde X(\sigma)=-\hat H Y(\sigma)$, where $\hat H=i\,\mbox{sign }\hat\kappa$ 
is the Hilbert operator. Such parametrization is used in the theory of surface waves 
at the deep water (see Refs.\cite{DKSZ96,DLZ95,RD2005PRE}, and references therein). 
For the present problem this choice has no special meaning, 
it is only important that arbitrary 2D curves can be represented in this way.
Then equations of motion for the two basic functions $Y(\sigma,t)$ and  $\psi(\sigma,t)$
can be written in a non-canonical form (for technical details, see Ref.\cite{RD2005PRE}),
\begin{eqnarray}
Y_t&=&\mbox{Im}\Big\{iZ'(1+i\hat H )\Big[{(\delta\tilde{\cal K}/\delta\psi)}/
{|Z'|^2}\Big]\Big\},\nonumber\\
\psi_t&=&{\mbox{Im}\Big\{(1-i\hat H)
\left[2({\delta\tilde{\cal K}}/{\delta Z})Z'
+ ({\delta\tilde{\cal K}}/{\delta\psi})\psi'\right]\Big\}}/{|Z'|^2}\nonumber\\
&&-\epsilon g\,\mbox{Im\,}Z
-\psi'\hat H\left[{(\delta\tilde{\cal K}/\delta\psi)}/{|Z'|^2}\right],
\end{eqnarray}
where $\psi'=\partial_\sigma\psi(\sigma,t)$, $Z'=\partial_\sigma Z(\sigma,t)$, and
\begin{equation}
Z(\sigma,t)=X(\sigma,t)+iY(\sigma,t)=\sigma+(i-\hat H)Y(\sigma,t),
\end{equation}
\begin{equation}
2(\delta\tilde{\cal K}/\delta Z)=(\delta\tilde{\cal K}/\delta \tilde X)
-i(\delta\tilde{\cal K}/\delta Y).
\end{equation}
With appropriate regularization at very short scales, the above equations 
can be efficiently simulated on computer using fast Fourier transform routines. 
It should be noted that for correct treatment of quite steep waves, 
say when $\kappa Y\sim A$ with $A\gtrsim 1$, 
one has to expand $\tilde F$ in Eq.(\ref{tilde_F}) 
up to a sufficiently high order  $N$ satisfying the condition ${(2A)^{N+1}}/{(N+1)!}\ll 1$. 
For example, the third-order approximation
Eq.(\ref{tilde_K}) can be good only up to $\kappa Y\approx 1$.

\emph{Summary and discussion. ---} To summarize, in this work a simple method has been
suggested for derivation of higher-order dispersive approximations in the theory
of fully nonlinear planar interfacial waves with a small density jump, 
in a constant-vorticity flow over non-uniform bed. Explicit quasi-local
expressions have been presented up to the 7th order.
The method is based on a variational formulation of the interface dynamics, 
and it uses an expansion of a Fourier transform of the Green's function 
entering a non-local part of the exact Hamiltonian. 

It should be noted that analogous expansion of a Fourier-transformed Green's function
can be used also in three-dimensional case for the deep-water purely potential regime.
Another possible application concerns the recently suggested two-layer compressible 
atmospheric model \cite{R2010JETP}. 
Details of the corresponding studies will be published elsewhere.

These investigations were supported by RFBR 
(grants 09-01-00631 and 07-01-92165),
by the ``Leading Scientific Schools of Russia'' grant 6885.2010.2,
and by the Program ``Fundamental Problems of Nonlinear Dynamics'' 
from the RAS Presidium.

\end{document}